\newif\ifsingle
\ifsingle
\documentclass[11pt,draftclsnofoot, onecolumn]{IEEEtran}		
\else		
\documentclass[journal]{IEEEtran}

\usepackage{times}
\usepackage{amsmath,dsfont}
\usepackage{amssymb,amsthm}
\usepackage{epsfig,verbatim}
\usepackage{subfigure}
\usepackage{setspace}
\usepackage{color}
\usepackage{cite}
\usepackage{epstopdf}
\usepackage{graphics}
\usepackage{accents}
\usepackage{acronym}
\usepackage[bookmarks,colorlinks]{hyperref}
\usepackage{booktabs}
\usepackage{mathtools}
\usepackage{algorithm}
\usepackage{algorithmic}
\usepackage{enumitem}

\definecolor{NewColor}{rgb}{0,0,0} %{0.2,0,0.5}

\ifsingle

\else

\fi % ------------------------------------------

\acrodef{adc}[ADC]{Analog-to-Digital Convertor}
\acrodef{cs}[CS]{Compressed Sensing}
\acrodef{dtft}[DTFT]{discrete-time Fourier transform}
\acrodef{dnn}[DNN]{deep neural network} 
\acrodef{csi}[CSI]{channel state information}
\acrodef{map}[MAP]{maximum a-posteriori probability}
\acrodef{snr}[SNR]{Signal-to-Noise Ratio}
\acrodef{bs}[BS]{Base Station} 
\acrodef{iot}[IOT]{Interent of Things}
\acrodef{mimo}[MIMO]{Multiple-Input Multiple-Output}
\acrodef{mmimo}[mMIMO]{massive Multiple-Input Multiple-Output}
\acrodef{mse}[MSE]{mean-squared error}
\acrodef{pdf}[PDF]{probability density function}
\acrodef{rv}[RV]{random variable}
\acrodef{fec}[FEC]{forward error correction}
\acrodef{rs}[RS]{Reed-Solomon}
\acrodef{lti}[LTI]{linear time-invariant}
\acrodef{wss}[WSS]{wide-sense stationary}
\acrodef{psd}[PSD]{power spectral density}
\acrodef{ser}[SER]{symbol error rate} 
\acrodef{ber}[BER]{bit error rate} 
\acrodef{isi}[ISI]{intersymbol interference}  
\acrodef{awgn}[AWGN]{additive white Gaussian noise} 
\acrodef{ut}[UTs]{User Terminals} 
\acrodef{mmw}[mmWave]{millimeter wave}
\acrodef{ris}[RIS]{reconfigurable intelligent surface} 
\acrodef{dma}[DMA]{Dynamic Metasurface Antenna} 

\IEEEoverridecommandlockouts

\title{Dynamic Metasurface Antennas for 6G\\  Extreme Massive MIMO Communications 
}
\author{
	\IEEEauthorblockN{Nir Shlezinger,~\IEEEmembership{Member,~IEEE,}  George C. Alexandropoulos,~\IEEEmembership{Senior Member,~IEEE,}  Mohammadreza F. Imani,~\IEEEmembership{Member,~IEEE,} Yonina C. Eldar,~\IEEEmembership{Fellow,~IEEE,} and David R. Smith,~\IEEEmembership{Senior Member,~IEEE}
	} 
	\thanks{
		N. Shlezinger  and Y. C. Eldar are with the Faculty of Math and CS, Weizmann Institute of Science, Rehovot, Israel (e-mail: nirshlezinger1@gmail.com; yonina@weizmann.ac.il). 	
	}
	\thanks{
		G. C. Alexandropoulos is with the Department of Informatics and Telecommunications, National and Kapodistrian University of Athens, 15784 Athens,	Greece (e-mail: alexandg@di.uoa.gr).
	}
	\thanks{
	 M. F. Imani and D. R. Smith are with the Department of ECE, Duke University, Durham, NC, USA (e-mail:  mohamad.imani@gmail.com; drsmith@duke.edu).
	}

	\vspace{-1.0cm}
	
}
\vspace{-0.75cm}

\begin{document}
	
	\maketitle
	\pagestyle{plain}
	\thispagestyle{plain}
		
	%----------------------------------------------------------------------------------------
	%	ABSTRACT
	%----------------------------------------------------------------------------------------
	\begin{abstract} 
		Next generation wireless base stations and access points will transmit and receive using extremely massive numbers of antennas. A promising  technology for realizing such massive arrays in a dynamically controllable and scalable manner with reduced cost and power consumption utilizes surfaces of radiating metamaterial elements, known as {\em metasurfaces}. To date, metasurfaces are mainly considered in the context of wireless communications as passive reflecting devices, aiding conventional transceivers in shaping the propagation environment. This article presents an alternative application of metasurfaces for wireless communications as active reconfigurable antennas with advanced analog signal processing capabilities for next generation transceivers. We review the main characteristics of metasurfaces used for radiation and reception, and analyze their main advantages as well as their effect on the ability to reliably communicate in wireless networks. As current studies unveil only a portion of the potential of metasurfaces, we  detail a list of exciting research and implementation challenges which arise from the application of metasurface antennas for wireless transceivers.
	\end{abstract}
	
	%----------------------------------------------------------------------------------------
	%	Introduction
	%----------------------------------------------------------------------------------------
	\vspace{-0.4cm}
	\section{Introduction}
	\vspace{-0.1cm}
	% Begin with the promising gains of massive MIMO systems.
The increasingly demanding objectives for 6th Generation (6G) communications have spurred recent research activities on novel transceiver hardware architectures and relevant communication algorithms. Such hardware architectures comprise large numbers of ElectroMagnetic (EM) radiating elements, thus paving the way for \ac{mmimo} communications. %It has been shown in \cite{marzetta2010noncooperative} that 
A \ac{mmimo} system with arbitrarily large number of antenna elements can provide substantial gains in spectral efficiency with relatively simple signal processing algorithms. This potential has motivated the incorporation of \ac{mmimo} technology in the 5G New Radio (NR) interface %\cite{B:Dahlman_5G_NR}, 
, and \ac{mmimo} transceivers with extremely large number of antennas are considered to continue being one of the key technologies for 6G communications \cite{saad2019vision}.

With the widespread deployment of Internet of Things (IoT) devices, % and the rapidly increasing number of smart end user equipment, 
the number of  nodes connected over wireless media is expected to reach the order of tens of billions in the next few years. To address these massive connectivity, high peak device rates and increased  throughput requirements, future wireless networks are expected to transit into dense deployments of coordinating extreme \ac{mmimo} transceivers (namely, \acp{bs} and access points), particularly in urban and indoor environments. While the theoretical gains of densely deployed \ac{mmimo} systems are still being unveiled \cite{bjornson2019massive},
implementing such systems in practice is a challenging task. In particular, realizing \acp{bs} with hundreds, or even thousands, of antenna elements being able to simultaneously serve multiple users, gives rise to a multitude of practical difficulties for conventional  sub-6 GHz bands as well  millimeter wave and THz bands. Among those challenges are  high fabrication cost, increased power consumption, constrained physical size and shape, and deployment limitations.

Over the last few years, metamaterials have emerged as a powerful technology with a broad range of applications, including wireless communications. Metamaterials comprise a class of artificial materials whose physical properties, and particularly their permittivity and permeability, can be engineered to exhibit various desired characteristics  \cite{smith2004metamaterials}. When deployed in planar structures (a.k.a. metasurfaces), their effective parameters can be tailored to realize a desired transformation on the transmitted, received, or impinging EM waves \cite{DSmith-2017PRA}. Such structures have been lately  envisioned  as a revolutionary means to transform any naturally passive wireless communication environment (the set of objects between a transmitter and a receiver constitute the wireless environment) to an active one \cite{huang2019reconfigurable,di2019smart}. %\cite{Liaskos_Visionary_2018,huang2019reconfigurable,di2019smart}. 
Their extremely small hardware footprint enables their cost-effective embedding in various 3D components of the environment (e.g., building facades and room walls/ceilings). 

\acp{dma} have been recently proposed as an efficient realization of massive antenna arrays for wireless communications \cite{Yoo2018TCOM, shlezinger2019dynamic}. They provide beam tailoring capabilities \textcolor{NewColor}{and facilitate processing of the transmitted and received signals in the analog domain in a flexible and dynamically configurable manner using simplified transceiver hardware. In addition, \ac{dma}-based architectures require much less power and cost compared with conventional antenna arrays (i.e., those based on patch arrays and phase shifters) eliminating the need for complicated corporate feed and / or active phase shifters.} \acp{dma} may comprise a large number of tunable metamaterial antenna elements that can be packed in small physical areas \cite{Akyildiz-2016NCN} for a wide range of operating frequencies. This feature makes them an appealing technology for the extreme \ac{mmimo} transceivers of 6G wireless networks. In contrast to passive metasurfaces that have received extensive attention recently \cite{huang2019reconfigurable,di2019smart}, the potential and capabilities of metasurfaces as active \ac{mmimo} antenna arrays, as well as their associated challenges and future research directions, have not yet been properly treated in the wireless communication literature.  

In this article, we discuss the promising application of \acp{dma} as \ac{mmimo} transceivers for future %ultra high data rate 
wireless communications. We commence with a brief introduction on metasurfaces considered  for wireless communications, highlighting the differences between their nearly passive and dynamic counterparts. We then present the unique properties of \acp{dma} as transceiving structures with Analog and Digital (A/D) processing capabilities. We  elaborate on the relationship between \acp{dma} and conventional hybrid A/D \ac{mmimo} architectures, which are based on phase shifters for analog processing \cite{ioushua2019family} and discuss 
%, their current applications in microwave imaging \cite{sleasman2017single} and radar systems \cite{sleasman2017experimental}, as well as their differences with passive metasurfaces. 
the main advantages and drawbacks of using \ac{dma}-based \acp{bs} for \ac{mmimo} communications. A simulation study evaluating the ability of \acp{dma} to reliably communicate \textcolor{NewColor}{with a reduced number of Radio Frequency (RF) chains by exploiting their inherent analog signal processing flexibility} is detailed. Next, we present experimental results demonstrating the strong potential benefits of \acp{dma} for wireless communication. We conclude the article with a description of open problems in this area of research and a discussion of future directions towards unveiling the potential of \acp{dma} for 6G wireless communications.

%----------------------------------------------------------------------------------------
%	Metasurfaces
%----------------------------------------------------------------------------------------
\section{Metasurfaces for Wireless Communications}
\label{sec:Metasurfaces}
\textcolor{NewColor}{
Metamaterials comprise a class of artificial materials in which macroscopic, structured elements mimic the atoms or molecules. Each metamaterial element behaves as an electric or magnetic polarizable dipole, the collection of which can often be characterized by an effective permittivity and permeability. These dipole moments can be engineered so as to achieve desired EM properties} \cite{smith2004metamaterials}. The underlying idea behind metamaterials is to introduce tailored inclusions in a host medium to emulate diverse targeted responses. 
\textcolor{NewColor}{
When metamaterial elements are distributed over a planar surface, they are often referred to as metasurfaces} \cite{HMIMOS}. Such 2-Dimensional (2D) structures support the individual tuning of each metamaterial element, allowing the metasurface to carry out different functionalities, such as radiation, reflection, beamforming, and reception of propagating waves \cite{DSmith-2017PRA}. \textcolor{NewColor}{The ability to stack a large number of elements in a limited surface area allows metasurfaces to achieve highly directed signalling enabling holographic beamforming \cite{bjornson2019massive}.} 
% We can consider also citing industrial pulications here \cite{black2016holographic,black2020methods}
The properties of each element can often be externally controlled, yielding a dynamic metasurface, also referred to as a reconfigurable intelligent surface \cite{huang2019reconfigurable}.
	
\begin{figure*}
	\centering
	\includegraphics[width = \linewidth]{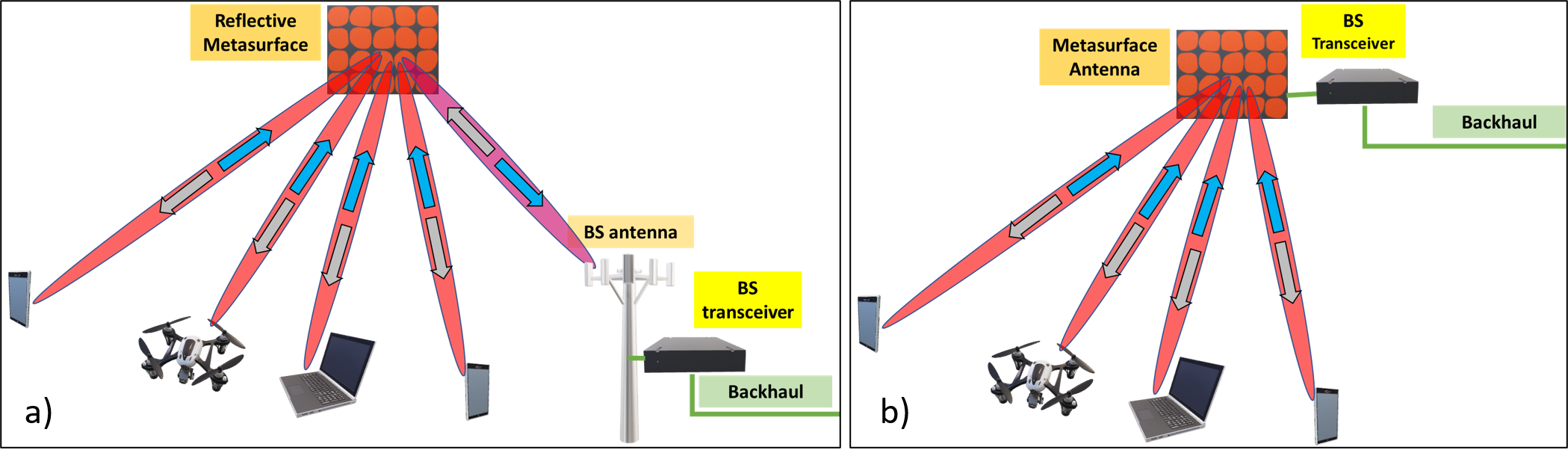} 
	\vspace{-0.6cm}
	\caption{Two applications of reconfigurable metasurfaces in the downlink (gray arrows) and uplink (blue arrows) of multi-user mMIMO wireless communications: a) as nearly passive reflective surfaces; and b) as active transceiver antenna arrays.
	}
%	\vspace{-0.4cm}
	\label{fig:PassiveVsActive1}
\end{figure*}

Two main applications of dynamically tuned metasurfaces have been considered recently in the context of wireless communications, as illustrated in Fig.~\ref{fig:PassiveVsActive1}. These are passive reflective surfaces \cite{ huang2019reconfigurable} and active transceiver antenna arrays \cite{shlezinger2019dynamic}. The former application builds upon the capability of metasurfaces to generate reconfigurable reflection patterns. In this case, metasurfaces deployed in urban settings can facilitate and improve  communication between the \ac{bs} and multiple users by effectively modifying EM signal propagation. \textcolor{NewColor}{The addition of a metasurface enables the communications system as a whole to overcome} harsh non-line-of-sight conditions and improve coverage, when the metasurface is placed within small distances from the BS or the users, without increasing transmission power. 
Nearly passive reconfigurable intelligent surfaces require some level of control to alter the impinging EM wave in light of the dynamic wireless environment. This is achieved by embedding a digital control unit, which is capable of tuning the metamaterial elements to obtain desired reflection patterns. Such reflective surfaces do not implement conventional relaying techniques (i.e., neither power amplification nor baseband processing) \cite{huang2019reconfigurable}, but only reflect the impinging signal in a controllable manner.

\acp{dma} comprise an additional recent usage of metasurfaces for communication as mMIMO antenna structures. This application exploits their ability to realize planar, compact, and low cost dynamically tunable massive antenna arrays \cite{shlezinger2019dynamic}, which can be deployed in current and future BSs and access points. As such, a DMA-based BS will consist of a multitude of radiating metamaterial elements that can transmit and receive communication signals over the wireless channel. By dynamically tuning  the EM properties of the DMA elements, one can control the analog beampattern for transmission and reception. Unlike passive reflective metasurfaces, which are  the focus of considerable research attention lately, the usage of metasurfaces as active mMIMO antennas is a relatively new area of research. To understand the potential of \acp{dma} for mMIMO systems, we discuss in the following section the architecture, properties, advantages, and drawbacks of using \acp{dma} for transmitting and receiving communication signals.

% TODO NIR CONTINUE READING FROM HERE APRIL 27 
 
%----------------------------------------------------------------------------------------
%	DMAs for comm
%----------------------------------------------------------------------------------------
%\vspace{-0.2cm}
\section{\acp{dma} for Massive MIMO Communications}
\label{sec:DMAs}
%\vspace{-0.1cm}
In this section, we discuss the application of metamaterial-based planar antenna arrays in wireless communication systems. We first detail the architecture of DMAs and their main characteristics in the context of communications, and then, elaborate on its operation when deployed for mMIMO \acp{bs}. Finally, some representative numerical evaluations are presented together with experimental results.
 
%-----------------------------------
%	Architecture and Model
%-----------------------------------
%\vspace{-0.2cm}
 \subsection{\ac{dma} Hardware Architecture}
 \label{subsec:DMAsModel}
\begin{figure*}
	\centering
	\includegraphics[width = \linewidth]{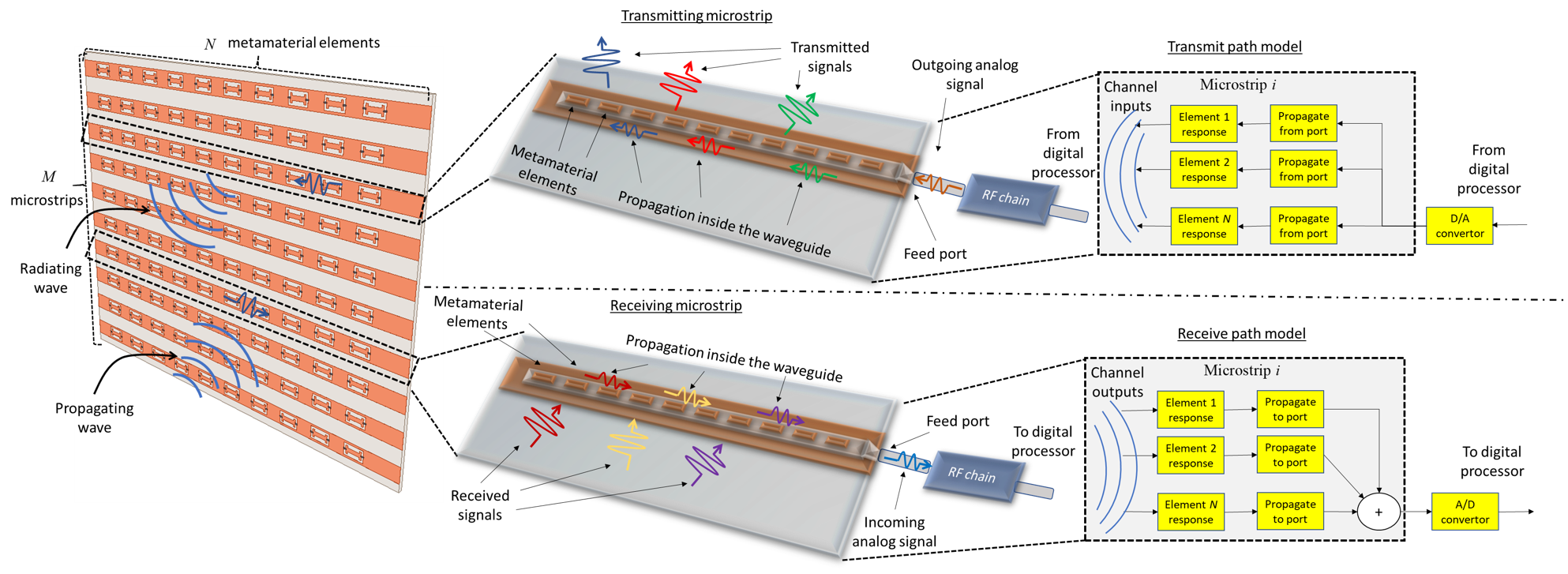} 
	\vspace{-0.6cm}
	\caption{An $N$-element \ac{dma} consisting of $M$ micropstrips, where each microstrip is implemented as an 1D waveguide. The upper right part and the lower right part of the figure illustrate their operation during transmission  and reception, respectively, along with their equivalent signal path models.  }
%	\vspace{-0.4cm}
	\label{fig:DMA_Diagram1}
\end{figure*}

%\vspace{-0.1cm}
 \acp{dma} consist of a multitude of reconfigurable metamaterial radiating elements that can be used both as transmit and receive antennas. Those elements are placed on a waveguide through which the signals to be transmitted, and the received waveforms intended for information decoding, are transferred. The transceiver digital processor, which generates the outgoing signals and processes the received signals, is connected to the waveguide through dedicated input and output ports, respectively. In general, \acp{dma} can use 2D waveguides connected to several input/output ports \cite{Yoo2018TCOM}. However, the common \ac{dma} architecture, on which we focus in this article, consists of multiple separate waveguide-fed element arrays, referred to as {\em 1-Dimensional (1D) waveguides}, each connected to a single input/output port, as illustrated in Fig. \ref{fig:DMA_Diagram1}. % in which microstrips are used as the waveguide of choice.  
 %
% \begin{figure}
%	\centering
%	\includegraphics[width = \columnwidth]{fig/Microstrip.png} 
	%\vspace{-0.8cm}
%	\caption{Illustration of a DMA microstrip used for the case of signal reception.}
%	\vspace{-0.4cm}
%	\label{fig:Microstrip}
%\end{figure}
% 
Such waveguides can accommodate a large number of radiating elements, which are commonly sub-wavelength spaced, allowing each input/output port to feed a multitude of possibly coupled radiators. Since this waveguide is typically designed to be single mode and the wave can only propagate along one line, its analysis is much easier than 2D waveguides (such as parallel plate waveguides), where a scattered wave from each element propagates in all directions. Furthermore, ensuring isolation between different ports is easier in 1D waveguides than in multiple ports of a 2D waveguide. A common implementation of 1D waveguides is based on \textit{microstrips}, as illustrated in Fig.~\ref{fig:DMA_Diagram1}. 

%The resulting two-dimensional \ac{dma} is a planar antenna obtained by stacking a set of one-dimensional waveguides. An illustration of a \ac{dma} with $M$ waveguides, each embedded with $N$ radiating unit elements, is depicted in Fig. \ref{fig:DMA_Diagram1}. In this schematic, microstrips are used as the waveguide of choice. 

When \acp{dma} are deployed as receive antennas, the signals captured at each metamaterial element propagate through the corresponding waveguide to the output port, where they are acquired and forwarded to the digital unit for baseband processing. 
%On the other hand, 
In a \ac{dma}-based transmit antenna array, the signals to be radiated from its metamaterial elements are fed to each waveguide via its input port.	The relationships among the radiating signals and those captured/fed at the input/output port of each waveguide are determined by the following two properties arising from the \ac{dma} architecture:
	\begin{enumerate}
		\item \label{itm:P1} Each metamaterial element acts as a resonant electrical circuit whose frequency response is described by a Lorentzian equation \cite{DSmith-2017PRA}. The parameters of the resonant circuit, i.e., its oscillator strength, damping factor, and resonance frequency, are externally controllable and can be configured in run-time for each element individually. An illustration of the normalized magnitude of the element response achieved for several different resonant frequencies is depicted in Fig.~\ref{fig:Lorentz1Mag}. This figure demonstrates that the elements can be tuned to exhibit diverse responses, varying from bandpass to frequency flat filters. 
		
		\begin{figure}
	    \centering 
		\scalebox{0.58}{\includegraphics{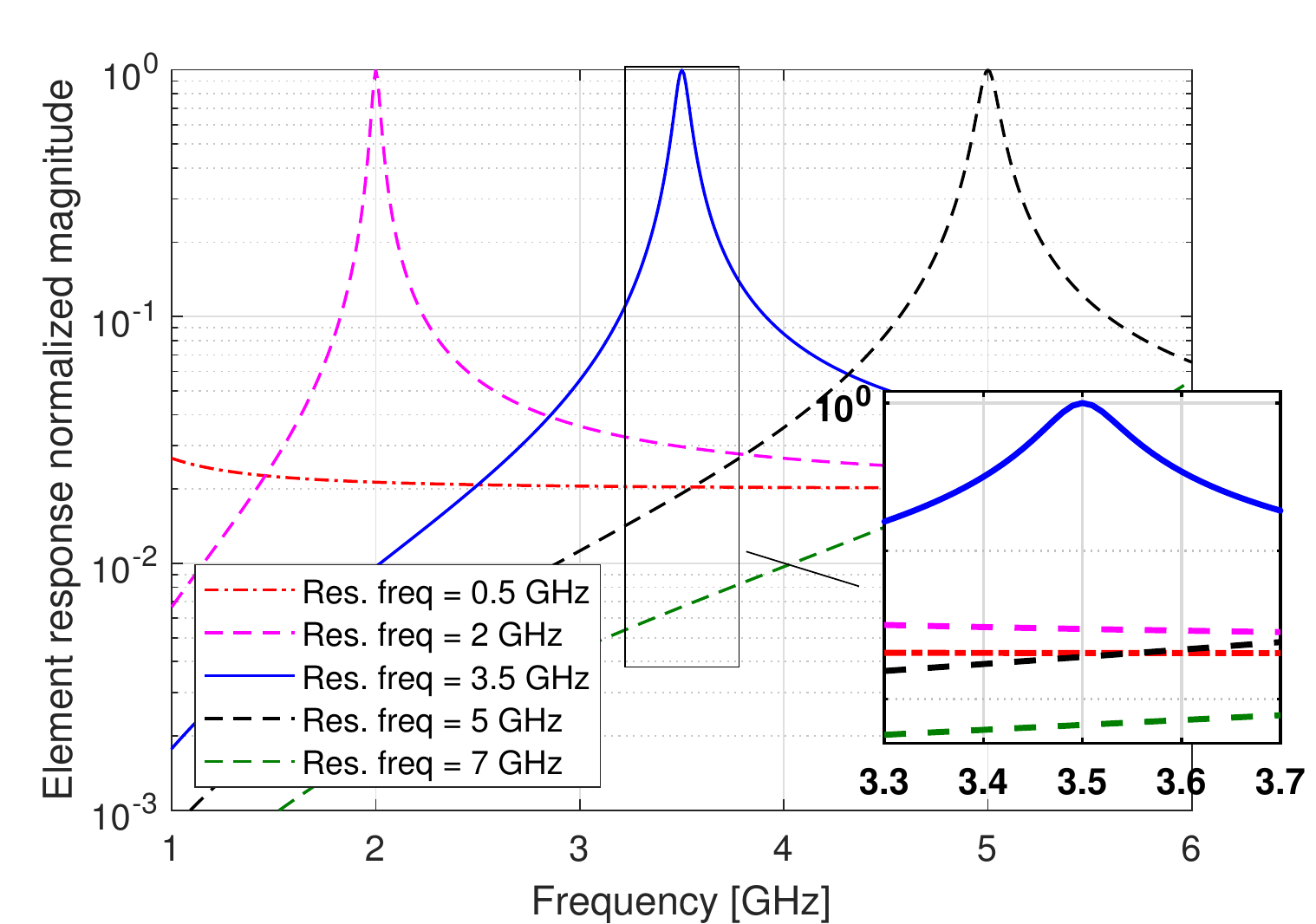}}
	%	\vspace{-0.2cm}
		\caption{The normalized element response magnitudes for different settings of the Lorentzian resonant frequency (Res. freq) as functions of the operating frequency.}
		\label{fig:Lorentz1Mag}		
	\end{figure} 	
		
		\item \label{itm:P2} In each waveguide, the signal has to travel between the feed port and each metamaterial element. Consequently, the signals propagating along the waveguide accumulate different phases for each metamaterial element. 
	\end{enumerate}
	An illustration of the resulting equivalent model for each DMA waveguide, which arises from the aforementioned properties, is depicted in the  right part of Fig.~\ref{fig:DMA_Diagram1}. As shown in this figure, the input/output port is connected to a set of radiating elements via a reconfigurable filter encapsulating the joint effect of the propagation inside the waveguide and the externally controllable frequency response of the metamaterial element.

%	\begin{figure}
%		\centering
%		\scalebox{0.48}{\includegraphics{fig/Lorentz1Phase.eps}}
%		\vspace{-0.2cm}
%		\caption{Element response phase vs. frequency.}
%		\label{fig:Lorentz1Phase}
%	\end{figure}

%-----------------------------------
%	Wireless Base Stations
%-----------------------------------
 \subsection{\ac{dma}-Based mMIMO Transceivers}
 \label{subsec:DMAsBSs}
Similar to conventional massive antenna arrays, \acp{dma} interface the electronic circuitry of a \ac{mmimo} transceiver with the EM waveforms propagating over the wireless channel. However, their application in \ac{mmimo} BSs induces several unique characteristics, which follow from the \ac{dma} structure discussed in the previous section. First, the number of independent data streams that can be processed by a \ac{dma}-based BS in the digital domain is typically much smaller than the number of its metamaterial antenna elements. In particular, a \ac{dma} consisting of $M$ waveguides (e.g., the 1D waveguides in Fig.~\ref{fig:DMA_Diagram1} termed as microstrips), each embedded with $L$  metamaterial antenna elements, allows the transceiver to process $M$ data streams, while utilizing in total $N=ML$ metamaterial elements. Consequently, \ac{dma}-based transceivers implement a form of hybrid A/D beamforming, since part of the processing of the transmitted and received signals is carried out in the analog domain, as an inherent byproduct of the waveguide-fed metamaterial array architecture. Such hybrid processing, which requires additional dedicated hardware in conventional \ac{mmimo} architectures, allows the \ac{bs} to utilize an amount of metamaterial elements that is much larger than the number of the digitally processed data streams. This inherent expansion upon transmission and compression in reception is typically desirable in \ac{mmimo} \acp{bs} as a method to reduce the number of costly RF chains \cite{ioushua2019family}, while achieving increased beamforming gain, as well as to facilitate efficient operation under low quantization requirements \cite{shlezinger2018hardware}.

In DMA-based transceivers, each metamaterial antenna element can exhibit a broad range of frequency responses with various combinations of amplitude and phase values, ranging from frequency selective to frequency flat profiles. These profiles are externally controllable for each element separately. This implies that \ac{dma}-based mMIMO \acp{bs} implement dynamically reconfigurable and frequency selective hybrid A/D beamforming by tuning their  elements to impart different levels of attenuation and phase shift on the transmitted and received signals. Consequently, \ac{dma}-based BSs can be treated as hybrid A/D beamforming systems which do not require additional dedicated analog combining circuitry, while offering increased flexibility, compared to conventional hybrid architectures consisting of interconnections of frequency static phase shifters and switches \cite{ioushua2019family}.

%The main advantages of using \acp{dma} for massive \ac{mimo} systems over utilizing large arrays of conventional antennas are implementation-oriented.

%as well as the family of manipulations that \acp{dma} can carry out in the analog domain, i.e., the analog precoding and combining which arise for the configuration of the metamaterial elements, which is unique for \acp{dma}.

Furthermore, DMAs provide similar beamforming capabilities to those achievable with typical phased array antennas, but with much lower power consumption and cost. To better illustrate this point, we note that the radiation pattern formed by a DMA is the superposition of the radiated field from many metamaterial radiators, whose complex amplitude are determined by two factors: the tunable resonance response of the elements and the phase accumulated by the guided wave. Using simple holographic techniques, the tuning states of the metamaterial elements can be designed to form beams in any direction of interest. Since the tuning of metamaterial elements is usually accomplished with simple components, such as varactors, it requires minimal additional power for beam steering. This is in contrast to many common antenna arrays which use active phase shifters that consume significant power. The beamforming capability of DMAs is heavily dependent on the subwavelength spacing of metamaterial elements. This is due to the fact that the metamaterial elements amplitude and phase cannot be tuned separately. To augment this limited design space, one can utilize the phase accumulation of the feed wave as it travels between different elements. A densely sampled waveguide thus provides enough degrees of freedom to form any desired beamforming pattern.

%\textcolor{red}{\bf [We should add here a discussion regarding the advantages of DMAs over patch arrays, e.g., in terms of beamforming gain. Reza - can you sketch a few sentences here on this topic?]}

 \subsection{Numerical and Experimental Results}
 \label{subsec:DMA_Prototype}
 The \ac{dma} architecture facilitates the implementation of massive amounts of metamaterial elements in simple, energy-efficient, compact, and low-profile design configurations, irrespective of the operating frequency. The planar physical shape of \acp{dma} makes them suitable for installations in urban and indoor environments, and the inherent expansion/compression induced by their waveguide-fed architecture reduces the number of required expensive RF chain circuits. %, resulting in a form of a hybrid A/D processing. 
However, this expansion/compression, which stems from the fact that the transceiver can access the signals only at each waveguide's input/output ports, reduces the  achievable performance compared to a fully-digital transceiver which is capable of simultaneously accessing the signals transmitted and received from each antenna element separately. In particular, this performance reduction follows from the DMA requirement for fewer digital streams than antenna elements, and is a common characteristic of hybrid A/D beamforming architectures. 

\begin{figure}
		\centering
		\scalebox{0.48}{\includegraphics{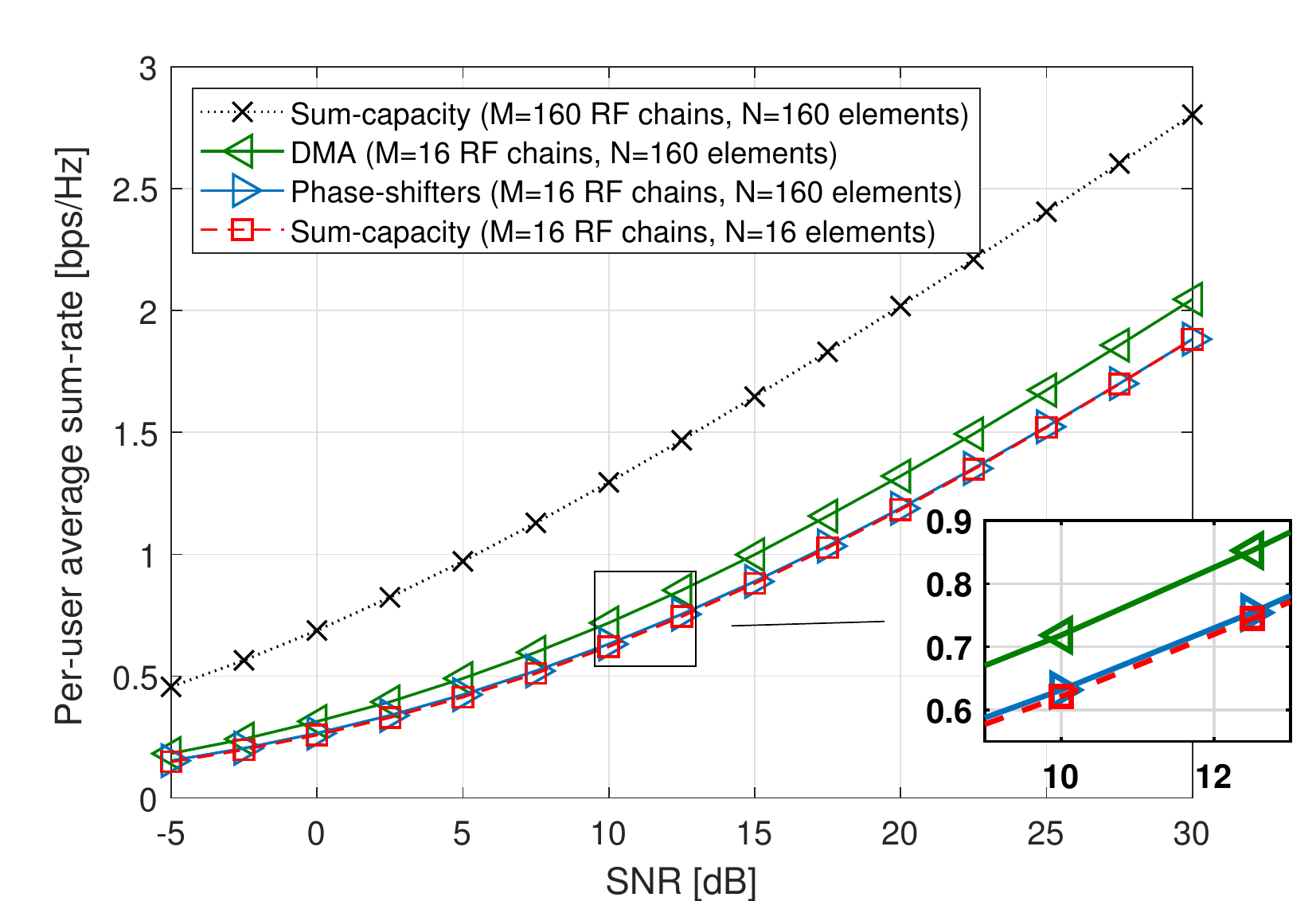}}
		%\vspace{-0.2cm}
		\caption{Achievable uplink sum-rate performance in bps/Hz versus the SNR in dB for a $160$-antenna mMIMO BS with $16$ RF chains serving simultaneously $64$ users within a cell with $400$m radius \textcolor{NewColor}{based on \cite{shlezinger2019dynamic}}. For the DMA architecture, each of the $M=16$ microstrips includes $L=10$ unit metamaterial elements. In the fully connected hybrid A/D beamforming architecture, the phase-shifter network interconnects each antenna element to all RF chains. }
		\label{fig:SEvsSNR_IFreq}
	\end{figure}

To evaluate the effect of the aforementioned properties on the performance of \ac{dma}-based \ac{mmimo} \acp{bs}, we depict in Fig.~\ref{fig:SEvsSNR_IFreq} the achievable uplink sum-rate performance of narrowband communications at $3.5$GHz carrier frequency as a function of the operating \ac{snr}. The achievable sum-rate is computed by treating the uplink wireless channel as a multiple access channel, and thus the number of RF chains is not restricted to be larger than the number of users.  % defined as one over the noise power.
In this figure, the same performance with a fully-connected hybrid A/D architecture is included, as well as the sum-capacity of the uplink channel when the number of \ac{bs} antennas equals the number $N$ of receiving elements and when it equals the number $M$ of receive RF chains. The corresponding simulation scenario, based on the setup detailed in \cite[Sec. IV]{shlezinger2019dynamic}, consists of $64$ user terminals uniformly distributed in a single cell of $400$m radius that simultaneously communicate in the uplink direction with a mMIMO \ac{bs} having $N=160$ antenna elements. For the DMA architecture, $L=10$ unit metamaterial elements are coated upon each of the $M=16$ waveguides (microstrips in this case). With the fully connected hybrid A/D beamforming architecture, all antenna elements are attached via phase-shifter networks to each of the $M=16$ RF chains. %\textbf{[Current mMIMO is at $3.5$GHz; can we get results for this frequency for this figure? Conventional hybrid beamforming is mainly discussed for mmWaves, eg 18 and 39 GHz in 5G. One another note, in this figure, we compare only with hybrid beamforming architecture with $M\ll N$, we could compare with conventional mMIMO too. For example, consider $M=N=64$ for a mMIMO at $3.5$GHz. We get the results for the typical cases, and for the DMA with keep $M=64$, but increase the $L$, thus $N$ and show the performance. The DMA provides increased beamforming gain], \textcolor{blue}{[The main motivation for including this figure is that the simulation setup is already detailed in \cite{shlezinger2019dynamic}. To present a new simulation setup would involve discussing all the details of the experiment. I do not think that this is a problem we can do that. More complicated is the comparison of DMAs to competing architectures with fully-digital architectures operating with the same number of RF chains. This follows since when comparing two receivers with different number of elements we are essentially observing different channels. Now we would have to think what do we wish to maintain fixed - the antenna aperture? its geometry? these selections would affect the resulting channel and hence the achievable rate. Anyway, I will try to do something here that we can justify. Also - is the rich scattering model used in \cite{marzetta2010noncooperative} also holds in the 3.5 GHz band? I guess so, and in this case I do not expect to see much difference as compared to the current figure]}}. 
As observed in Fig.~\ref{fig:SEvsSNR_IFreq}, the DMA architecture yields sum-rate performance closer to the sum-capacity with $M=N$ than conventional hybrid A/D architectures based on fully-connected networks of phase shifters. This advantage follows from the previously discussed capability of \acp{dma} to implement different forms of analog signal processing. The  rate gains of \acp{dma} with $M$ RF chains and $N=ML$ elements over the sum-capacity when the \ac{bs} has $M$ antennas indicates that, connecting each RF chain to a microstrip with multiple elements can notably improve the communication rate compared to using each RF chain to feed a single antenna. Similar results have been reported for the downlink case %in \cite{wang2019dynamic}, and in \cite{wang2020dynamic} 
as well as
for communications with low-resolution analog-to-digital converters \cite{wang2020dynamic}. %This demonstrates that the advantages of using \acp{dma} are not purely implementation-oriented, but can also be translated into performance gains.
 
\begin{figure}
		\centering
		\scalebox{0.5}{\includegraphics{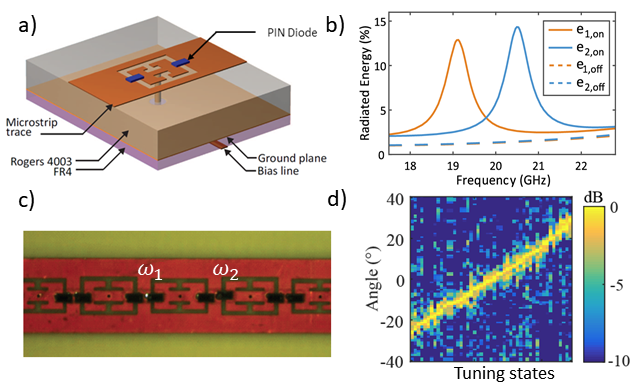}}
		%\vspace{-0.2cm}
		\caption{\textcolor{NewColor}{An experimental implementation of a dynamic 1D waveguide-fed metasurface \cite{sleasman2017experimental,sleasman2016design}:} a) Detailed circuitry of a reconfigurable metamaterial element. b) The simulated response of the metamaterial element, where the impact of the PIN diodes to render the element in radiating ($e_{1,\rm on}$ and $e_{2,\rm on}$) and non-radiating ($e_{1,\rm off}$ and $e_{2,\rm off}$) states are shown. c) A close-up view of a sample 1D DMA with metamaterial elements having two different resonance frequencies. d) Beamforming capability of the fabricated 1D DMA.}
		\label{fig:DMA_sample}
	\end{figure}

Despite the growing interest in using reconfigurable metasurfaces (either as an intelligent reflector or a transceiver) to augment wireless communication systems, very few experimental studies have been conducted. Most experimental studies focus on verifying the possibility to model each tunable metamaterial radiator as a polarizable dipole and, optimize their tuning states to form desired beams. For instance, the \ac{dma} configuration illustrated in Fig.~\ref{fig:DMA_Diagram1} consists of an array of microstrip, each comprised of a 1D waveguide-fed metasurface. An experimental implementation of such a dynamic 1D waveguide-fed metasurface, based on \cite{sleasman2017experimental,sleasman2016design}, is shown in Fig.~\ref{fig:DMA_sample}. This waveguide is fashioned with metamaterial elements with two different (interleaved) resonant frequencies. Each element is loaded with two PIN diodes, which render the metamaterial element radiating and non-radiating, as shown in Fig.~\ref{fig:DMA_sample}. A close-up view of a fabricated metamaterial is also shown. In addition, the figure includes experimental results for the ability of the metasurfaces to form beams in different directions, showcasing the potential of DMA-based mMIMO transceivers in forming multiple directed beams. %It is worth emphasizing this demonstration is primarily geared toward imaging/radar systems. 

%\textcolor{red}{Discuss the prototype and the results. We are allowed to add two more figure for this purpose: one for the prototype and one for a performance experiment (if needed a figure can have multiple subfigures). {\bf We are allowed up to 6 figures in total for the entire article}.}

%----------------------------------------------------------------------------------------
%	The Road Ahead and Future Challenges
%----------------------------------------------------------------------------------------
\section{Open Research Challenges}
\label{sec:Challenges}
As previously discussed, the hardware architecture of \acp{dma} enables  efficient implementation of extreme mMIMO transceivers, which renders this concept a strong candidate technology for the physical layer of 6G communications. However, \acp{dma} come with certain design challenges that need to be carefully addressed, while giving rise to new opportunities. In the following, we list some of the most important related research challenges to date.

\textbf{Frequency-Selective Analog Beamforming:} Current algorithmic designs for \acp{dma} focus either on narrowband communications or ignore their capability to dynamically configure the frequency-selective profile of each unit metamaterial element. This unique property, which does not exist in any conventional hybrid A/D architecture, provides increased flexibility for the design of wideband operation by matching the spectral behavior of each element to optimize the equivalent wideband channel. Consequently, the true potential of \ac{dma}-based mMIMO systems in achieving reliable and ultra high rate communications is not yet fully explored. 

\textbf{Wireless Channel Estimation and Tracking:} To date, studies on \ac{dma}-based communications assume that the transceiver has full channel knowledge. In practice, however, the channel coefficients need to be efficiently estimated, which is a challenging task with hybrid A/D receivers. When channel estimation is carried out in a time-division duplexing manner, \acp{dma} offer the possibility of tuning their elements to facilitate channel estimation via pilot signals, and to adapt in a manner which optimizes data reception in light of the estimated channel. The design and analysis of efficient algorithms for \ac{dma}-based \acp{bs}, which have to estimate features of the wireless channel and reliably communicate have not yet been properly treated in the literature. 

\textbf{Hardware Design and Experimentation:} A large body of fabricated designs and experimental work is still required in order to transit the DMA concept into an established technology for future extreme mMIMO \acp{bs} with hundreds or thousands of metamaterial elements. The experimental studies detailed in the previous section  are an important first step which demonstrates the feasibility of this concept. Yet, the performance and other implications of using \ac{dma}-based \acp{bs} need be tested in a broad range of wireless setups. To date, most experimental studies of \acp{dma} are geared toward imaging and radar systems.

In particular, several key properties of \acp{dma} should be thoroughly tested to fully understand their potential for mMIMO communications. First, one  needs to quantify the level of correlation among the signals received by different metamaterial elements on the same waveguide as well as different waveguides. The impact of this correlation on the overall response needs to be investigated. In the example discussed in the previous section, the tuning of the metamaterial elements is optimized based on certain communication criteria. However, the resulting pattern may introduce more correlation or result in low \ac{snr} (for example, in a scenario where the resulting pattern is directed toward directions other than the intended users). The co-design of the metamaterial elements tuning states with the resulting pattern is another important factor that needs to be examined. When used as a transmitter, the possibility of the high amplitude RF signal (the carrier) to push the tunable component of each metamaterial element into a nonlinear regime needs to be investigated. This nonlinearity can cause intersymbol or intercarrier interference. 
Finally, although conceptually feasible, \acp{dma} for communications in the high millimeter wave (around 60 GHz) and THz (above 100 GHz) bands are still in relatively early stages of research.
%Examining DMAs from aspects that are of interest from wireless communication perspective should be a primary future direction.

\textbf{Hybrid Passive and Active Metasurfaces:}
As discussed in the introduction, the emerging concept of reconfigurable metasurfaces, with both its passive and active counterparts, is lately gaining increased interest for both controllable reflection and transmission/reception. It is reasonable to envision hybrid passive and active reconfigurable metasurfaces, notably strengthening the design flexibility for such surfaces and their performance gains in programmable wireless environments. For instance, having such a hybrid metasurface acting as a receiving device can significantly facilitate channel estimation \cite{GA2020_RIS}, which is currently a major challenge and a source of substantial overhead in purely passive metasurfaces. In addition, hybrid metasurfaces will enable more advanced relaying strategies, overcoming the dominating impact of pathloss in their passive versions. 

\textbf{Use Cases and Applications:} The use cases and applications where \acp{dma} can provide substantial improvement compared to current architectures have not yet been thoroughly identified. For example, their planar shape and reduced size for a small number of RF chains facilitate their deployment in indoor environments, like buildings, malls, train stations, and airports. In such setups,  \acp{dma} are expected to communicate with multiple users in close line-of-sight conditions, possibly operating in the near-field regime. Such near-field scenarios bring forth the possibility of beam focusing, 
%\cite{GA2020_TR}, 
namely, the ability to focus the signal towards a specific location, instead of a specific direction as in far-field conditions via conventional beamforming, allowing multiple orthogonal links to exist in the same indoor or outdoor area. % \cite{nepa2017near}. 

%----------------------------------------------------------------------------------------
%	Conclusion
%----------------------------------------------------------------------------------------
\section{Conclusion}
\label{sec:Conclusion}
\acp{dma} are an attractive radiating technology for next generation wireless systems, allowing to realize dynamically controllable \ac{mmimo} antennas of reduced cost and power consumption compared to conventional arrays. In this article we surveyed some of the key properties of \acp{dma} in the context of \ac{mmimo} systems, including their operation model during their transmission and reception, tunable frequency selective profile, beam steering capabilities, as well as advantages and drawbacks compared to conventional arrays. We concluded with a set of key open research directions, which are expected to pave the way in unveiling the full potential of using active metasurfaces in 6G wireless communications. 

%\textcolor{red}{\bf [We are allowed up to 15 references] }
	
%----------------------------------------------------------------------------------------
%	BIBLIOGRAPHY
%----------------------------------------------------------------------------------------
	\bibliographystyle{IEEEtran}
	\bibliography{IEEEabrv,refs}

\end{document}